# RAFDA: Middleware Supporting the Separation of Application Logic from Distribution Policy

Scott M. Walker, Alan Dearle, Stuart J. Norcross, Graham N. C. Kirby, Andrew J. McCarthy

*School of Computer Science*
*University of St Andrews*
*St Andrews*
{scott,al,stuart,graham,ajm}@dcs.st-and.ac.uk

**Abstract.** Middleware technologies, often limit the way in which object classes may be used in distributed applications due to the fixed distribution policies imposed by the Middleware system. These policies permeate the applications developed using them and force an unnatural encoding of application level semantics. For example, the application programmer has no direct control over inter-address-space parameter passing semantics since it is fixed by the application's distribution topology which is dictated early in the design cycle by the Middleware. This creates applications that are brittle with respect to changes in the way in which the applications are distributed. This paper explores technology permitting arbitrary objects in an application to be dynamically exposed for remote access. Using this, the application can be written without concern for its distribution with object placement and distribution boundaries decided late in the design cycle and even dynamically. Inter-address-space parameter passing semantics may also be decided independently of object implementation and at varying times in the design cycle, again, possibly as late as run-time. Furthermore, transmission policy may be defined on a per-class, per-method or per-parameter basis maximizing plasticity. This flexibility is of utility in the development of new distributed applications and the creation of management and monitoring infrastructures for existing applications.

## Introduction

Existing middleware systems including CORBA [1], Java RMI [2], Microsoft DCOM [3] and Microsoft .NET Remoting [4] suffer from several limitations that restrict the kinds of application that can be created using them and hamper their flexibility with respect to distribution and adaptability. In this paper we focus on four of these limitations, namely,

1. They force decisions to be made early in the design process about which classes may participate in inter-address-space communication.
2. They are brittle with respect to changes in the way in which the applications are distributed.

3. It is difficult to understand and maintain distributed applications since the use of middleware systems may force an unnatural encoding of application level semantics.
4. It is difficult to control the policy used to determine how objects are transmitted among the available address-spaces in a distributed application.

**Early Design Decisions** - The industry standard middleware systems all require the programmer to decide at application design time which classes will support remote access and to follow similar steps in order to create the remotely accessible classes. The programmer must decide the interfaces between distribution boundaries statically then determine which classes will implement these interfaces and thus be remotely accessible. These classes, known as *remote classes*, are hard-coded at the source level to support remote accessibility and only instances of these classes can be accessed from another address-space. Therefore, the programmer must know how the application objects will be distributed at run-time before creating any classes.

These middleware systems require the manual creation of ancillary code such as skeletons, proxies and stub implementation classes, which must extend special classes, implement special interfaces or handle distribution related error conditions, based on programmer-defined interfaces. All require the creation of server applications that configure the middleware infrastructure then instantiate and register objects for remote access.

**Brittleness with Respect to Change -** The brittleness of distributed applications created using existing middleware systems is due to the fact that the distribution of the application must be known early in the design process. The possible partitions of a distributed application are dependent on which classes within the application are remotely accessible and so the classes of object that can be separated from their reference holders is restricted. The problem of brittleness and inflexibility to change is more than a question of support for remote accessibility within application classes.

**Distorted Application Level Semantics** - Industry standard middleware systems decide the parameter-passing semantics applied during remote method call statically based on the remote accessibility of the application classes. In general, remotely accessible objects are passed by-reference and other objects are passed-by-value, though CORBA exhibits the same limitation in a slightly different way; only CORBA components may be transmitted across the network and each class is explicitly defined as either pass by-reference or pass by-value.

The parameter-passing semantics is tightly bound to the distribution of the application so changes to the distribution of an application have the side effect that application semantics may be altered. All objects of the same class must be transmitted in the same way, whether this is appropriate or not, and the programmer does not have the freedom to choose different parameter-passing semantics for classes on a per-application or per-call basis.

Since industry standard middleware systems force remotely accessible classes to extend special classes, implement special interfaces or handle network related errors explicitly, it is not possible to make application classes remotely accessible unless their super-classes also meet the necessary requirements. At best, this forces an unnatural or inappropriate encoding of the application semantics because classes are forced to be remotely accessible for the benefit of their sub-classes and, at worst, application classes that extend library classes cannot be remotely accessible at all.

This paper introduces RAFDA [5] a Java Middleware that permits arbitrary application objects to be dynamically exposed for remote access. Object instances are exposed as Web Services through which remote method invocations may be made. RAFDA has four notable features that differentiate it from other Middleware technologies.

1. The programmer does not need to decide statically which component classes support remote access. Any object instance from any application, including compiled classes and library classes, can be deployed as a Web Service without the need to access or alter the application's source code.
2. The system integrates the notions of Web Services, Grid Services and Distributed Object Models by providing a remote reference scheme, synergistic with standard Web Services infrastructure extending the pass by-value semantics provided by Web Services with pass by-reference semantics. Specific object instances rather object classes are deployed as Web Services further integrating the Web Service and Distributed Object Models. This contrasts with systems such as Apache Axis [6] in which object classes are deployed as Web Services.
3. Parameter passing mechanisms are flexible and may be dynamically controlled through policies. A deployed component can be called using either pass by-reference or pass by-value semantics on a per-call basis.
4. The system automatically deploys referenced objects on demand. Thus an object *b* that is returned by method *m* of deployed object *a* is automatically deployed before method *m* returns.

The process of implementing the application logic is thus separated from the process of distributing the application. Since any object can be made remotely accessible, changes to distribution boundaries do not require re-engineering of the application, making it easier to change the application's distribution topology. This separation of concerns simplifies the software engineering process to the programmer's advantage both when creating a distributed application and introducing distribution into an existing application. This simplifies the creation of tools such as monitoring and management components that need to access and modify object state from outwith those objects' local address space. Using traditional middleware systems, it is difficult to attach such tools to existing objects without access to source code and extensive engineering effort.

This functionality is provided by the RAFDA Run-Time (RRT), a Middleware system for Java development that tackles the problems inherent in existing middleware systems. The RRT simplifies the kinds of tasks that are common to the creation of distributed application such as dynamically exposing objects for remote access, obtaining remote references to remotely accessible objects and remote method invocation. The RRT can ensure the preservation of local application semantics in a distributed application and can automate object placement based on programmer-defined policies. Although the RRT is written in Java and is designed to support Java, it does not however employ any language-specific features unique to Java and so the techniques described here are applicable in other languages.

## Exposing Arbitrary Objects for Remote Access

The RRT permits arbitrary application objects to be exposed for remote access. Specific application objects rather than application classes are exposed via Web Services. In order to make an object remotely accessible it is first *deployed*; that is registered with the RRT, which exposes the object to remote access. Deployment creates a Web Service running within the RRT that uses the deployed object as the underlying *service object* on which incoming Web Service requests are performed. In effect, the RRT maps Web Service requests to method calls on object instances and performs appropriate encoding of the results. Deployed objects may be referenced by other local objects and neither the reference holders nor the deployed objects are aware of the deployment.

```
public static void deploy(Object objectForDeployment,
   Class deploymentInterface,
   String serviceName) throws Exception;
```

**Figure 1: The *deploy()* method.**

The signature of the *deploy()* method is shown in Figure 1. This takes three parameters which specify the object to be deployed, the interface with which the object is to be deployed, and a logical name for accessing the object. A number of issues arise from this simple method. Firstly, the *objectForDeployment* need not implement any special interfaces or extend any particular classes, maximizing flexibility. Secondly, the *objectForDeployment* need not implement the interface specified in the *deploymentInterface* parameter although it must be structurally compliant with that interface. This again maximizes flexibility and permits classes to be remotely exposed even if they were not envisioned to be so at design time. The *deploymentInterface* parameter can be a class or an interface but in either case the method signatures are extracted to form the Web Service interface for the deployed object. The *deploymentInterface* parameter is optional and if omitted, the object is deployed with an interface matching its concrete type. The *deploymentInterface* is a mechanism to allow control over which methods may be called remotely on an object and is supplied on a per-object, not a per-class, basis. Any method can be made remotely accessible, irrespective of its local protection modifier. The *deploymentInterface* acts as a remote protection mechanism that is independent of the local protection mechanism of the implementation language; in Java, the *public*, *protected*, *private* and default modifiers. Only the methods listed in the deployment interface are remotely accessible and, by default, the RRT will deploy only the public methods. The *servicename* parameter, which is also optional, permits the deployed object to be addressed using a logical name which must, of course, be unique within the deploying address space. Deployment can fail, resulting in an exception, if the deployment interface contains methods that do not exist in the class of the object being deployment or if the specified service name is already in use.

The *deploy* method may be called multiple times with the same *objectForDeployment* parameter with different *deploymentInterface* and *serviceName* parameters. This allows the programmer to expose the object with different logical names and potentially different interfaces.

An object of any class can be deployed including precompiled classes and those with native members. There is one caveat: the Web Services model provides no facility to allow field access, only method call. Thus the fields of a deployed object cannot be directly accessed and if the object does not provide *get()* and *set()* accessor methods then the fields cannot be accessed at all. This is a problem for all Java Middleware systems since field access cannot be intercepted. To address this problem, the RRT generates named *accessor* methods automatically at deployment time and adds them to the Web Service interface for the deployed object.

To illustrate the use of *deploy*, we use a small Peer-to-Peer (P2P) application as an example. A programmer has implemented a class called *P2PNode* which represents a node in a P2P routing network. This class is shown in Figure 2. This class has not be written with distribution in mind and does not implement any special interface or extend any base classes.

```
public class P2PNode {
   private final Key key;
   public P2PNode(Key key){…}
   public void addPeer(P2PNode peer){…}
   public void route(Key key, Message msg){…}
   public String getLog(){…}
   public void stop(){…}
   public void start(){…}
}
```

**Figure 2: The P2P Node Implementation**

Figure 3 shows how another programmer could deploy an instance of this class as part of some P2P application. The programmer wishes to expose the functionality of the node using three different interfaces — a management interface for controlling the node remotely, a monitoring interface and an interface exposing the P2P functionality. These interfaces are named *IManage* , *IMonitor* and *IP2PNode* respectively. Each of these interfaces is associated with the names *Manage*, *Monitor* and *P2P* respectively. It is assumed that these are well known names that are used by client programmers to access the services.

```
public interface IManage {
   public void stop();
   public void start();
}
public interface IMonitor {
   public String getLog();
}
public interface IP2PNode {
   public void addPeer(IP2PNode peer);
   public void route(Key key, Message msg);
   public Key getKey();
}
```

```
public class DeployP2PApp {
   public static void main(String[] args){
      P2PNode p2pNode = P2PAppFactory.makeNode()
      RAFDARunTime.deploy( p2pNode, IManage.class,
           "Manage" );
      RAFDARunTime.deploy( p2pNode,
           IMonitor.class, "Monitor" );

      RAFDARunTime.deploy( p2pNode,
           IP2PNode.class, "P2P" );
   }
}
```

**Figure 3: Deploying an instance of P2P Node**

### Browsing exposed objects

Deployed objects may be accessed either using their service name or a Globally Unique Identifier (GUID) allocated to the service at deployment time. Both of these may be discovered dynamically by clients. Typically, an application will deploy a small collection of objects with well known names thus often avoiding the need for dynamic GUID discovery. Deployed objects may addressed using a URL of the following form:

```
http://<machineName>:<port>/<NAME or GUID>
```

The RRT contains a web server and provides a web interface that can be accessed using a conventional web browser to obtain information about deployed objects. Each deployed service is listed, showing the deployment interface, service name, a string representation of the service object and a link to the WSDL.

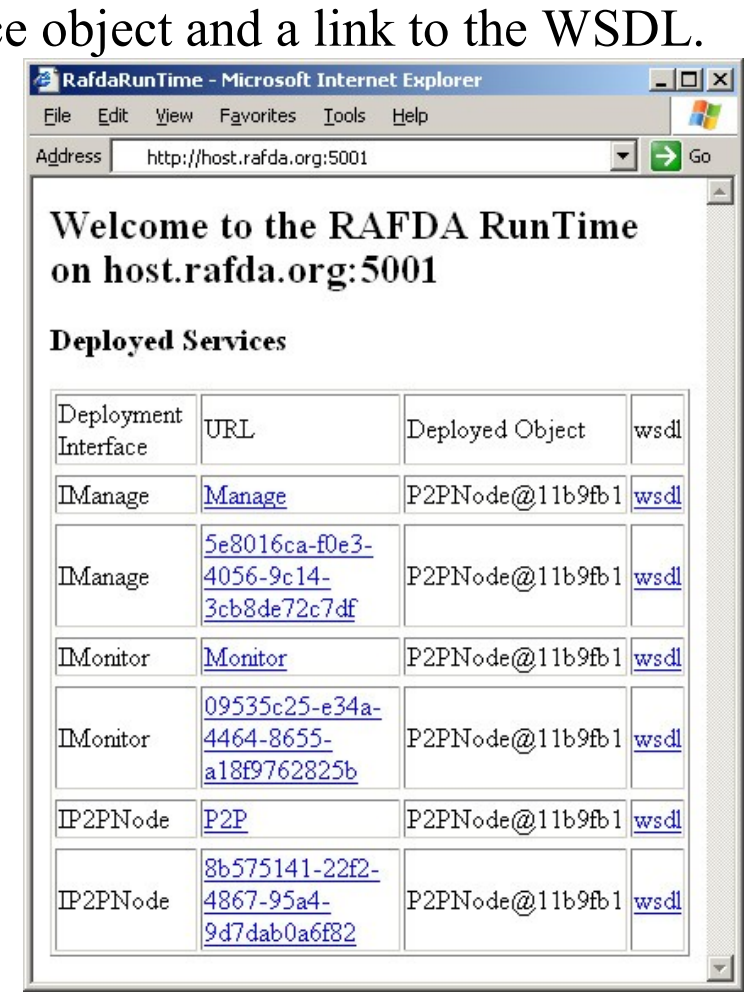

**Figure 4. Browsing an RRT**

Since a deployed object appears to remote clients as if it were a normal Web Service, the RRT can be used as a Web Services container. Like conventional Web Services containers, a list of available services and the WSDL for a particular service can be obtained from the RRT. Since WSDL is used to describe the methods provided by each service in a standardized manner, deployed objects are accessible using any Web Services technology, not just RRT-based clients. Figure 4 shows an RRT being browsed after the deployment code shown in Figure 3 has been executed.

## Client-side Distributed Object Programming using the RRT

The RRT may be used by client-side programmers to access remote objects. The RRT provides a method called *getObjectByName()* that permits a handle to be obtained to a deployed object. As will show later, the handle returned may be a reference to a proxy for a remote object, a local copy of the object or a hybrid of the two (a smart proxy). The *getObjectByName*() method takes three arguments as shown in Figure 5. These identify the *host* name of the machine on which the remote RRT runs, the *port* to which it is connected and the *name* with which the requisite object was deployed. The *name* can be either the programmer-defined service name or the automatically generated object GUID.

```
public static Object getObjectByName(String host,
     int port, String name) throws Exception;
```

**Figure 5: The *getObjectByName()* Method**

The object returned by *getObjectByName*() is same type as the deployment interface, to which it may be cast. Figure 6 shows the client-side code necessary to use the *P2PNode* deployed in Figure 3. The object returned by *getObjectByName()* is cast to type *IP2PNode* which was the interface used to specify its deployed type.

```
public class P2PClient {
   private String node = "host.RAFDA.org";
   private int port = 5001;

   public void deliver(Key dest, Message msg )
        throws Exception {
     IP2PNode node = (IP2PNode)
     RAFDARunTime.getObjectByName( node, port, "P2P" );
     node.route( dest, msg );
   }
}
```

**Figure 6: Client side code accessing a remote *P2PNode***

### Failure

Distributing an application introduces new failure modes. The RRT treats network failure differently from application failure. The RRT propagates application exceptions across the network and throws them locally for the client application to catch and handle. By contrast, network failures may either be hidden from or handled by the application programmer. The programmer indicates if network failures are to be handled buy the application by adding a *throws java.rmi.RemoteException* clause to the appropriate interface methods. If such a clause is present, the RRT will propagate network failure exceptions to the client, otherwise they are handled by the RRT. By default the RRT will continue to execute if possible, log the exception and returning null or zero values as results to the remote method calls. However it can be configured to exhibit fast-fail behaviour in the event of exceptions.

## Controlling Object Transmission Policy

As described in the introduction, using traditional middleware, the distribution topology of an application determines the object transmission semantics that are employed during remote method calls. For example, in Java RMI [2], only classes that implement the *java.rmi.Remote* interface and meet certain other criteria may be deployed for remote access. Such objects are always passed by-reference if they are accessed across an address space boundary. All other objects that traverse address-space boundaries must be instances of classes that implement the *java.io.Serializable* interface and these objects are always passed by-value. This problem can also be observed in Microsoft .NET Remoting [4], CORBA [1] and Web Services [7].

Within a single application, instances of some class may be required to be transmitted by-value or by-reference depending on the circumstances. In most existing middleware systems this would require that different classes be created. Hybridisation is sometimes desirable whereby some object state is cached at a client whilst other state is remotely accessed. Using the RRT's transmission policy framework, the application programmer can employ the most advantageous object transmission policy for the circumstances. In addition to providing the programmer with the flexibility to control the application semantics, the dynamic specification of policy independently of class implementation allows the roles of library class programmer and application programmer to be separated. The library class programmer is concerned only with the functional requirements. Thus, library classes make fewer assumptions about the environment in which they are to be deployed and the application programmer has the freedom to apply any parameter-passing policy to instances of any class, increasing the likelihood that any given class will be reusable in another context.

### Defining Transmission Policy

By default the RRT passes objects by-reference when interacting with other RRTs and by-value when interacting with standard Web Service Clients. However, the transmission policy framework described here provides a mechanism to allow the programmer to dynamically specify how objects should be transmitted during inter-RRT remote method calls. This is achieved using the local RRT's *TransmissionPolicyManager* which supports five types of policy rule as shown in Figure 7 and contains query methods (not shown) which permit the policies that are currently in place to be inspected. Each of type of policy rule has an associated *set* and *get* (again not shown) method.

```
public class TransmissionPolicyManager {
   public static void setMethodPolicy(String className,
        String methodName, int policy, int depth,
        boolean isOverrideable) { ... }

   public static void setReturnValuePolicy(
        String className, String methodName, int policy,
        boolean isOverrideable) { ... }

   public static void setParamPolicy (String className,
        String methodName, int paramNumber, int policy,
        int depth, boolean isOverrideable) { ... }

   public static void setClassPolicy(String className,
        int policy, boolean isOverrideable) { ... }

   public static void setFieldToBeCached(
        String className, String fieldName) { ... }
}
```

**Figure 7: The TransmissionPolicyManager**

The five types of rule are as follows:

- *Method policy rules* are associated with methods as a whole and are set using the *setMethodPolicy()* method. This method specifies how method arguments should be transmitted. For example, a method policy rule might specify that during a call to a particular method, the arguments should all be passed by-reference. The parameters to *setMethodPolicy* include the method name to which the policy applies, the policy to be applied (using static values from *PolicyType* shown below), the depth to which the closure of the parameters should be traversed in the case of pass by-value, and whether the policy may be overridden (discussed below).
- *Return policy rules*, set using the *setReturnPolicy()* method, are also associated with methods but control how the return values from methods should be transmitted. For example, a return policy rule might specify that the return value from a particular method should be passed by-value. The method policy rule and return policy rule associated with a single method are independent of each other and need not specify the same behaviour. The *setReturnPolicy()* method takes the same arguments as the *setMethodPolicy()* method which apply to the return value rather than the parameters.
- *Argument policy rules*, set using the *setArgumentPolicy()* method, are associated with individual method arguments and indicate how particular arguments within a method signature should be transmitted. They allow the programmer fine-grain control over the policy that is applied to each of the arguments of a method. The parameters to this method are similar to the previous two but an extra parameter is required to specify the parameter to which the policy applies.

- *Class policy rules*, set using the *setClassPolicy()* method, are associated with classes rather than methods and indicate how instances of particular classes should be transmitted. For example, a class policy rule might specify that all instances of a particular class should be passed by-value. Class policy rules are applied based on the actual classes of the transmitted objects, rather than the classes specified in the method signature, which may be super-classes of the arguments. Class policy rules do not take a depth parameter since the object classes they reference may have a class policy associated with them. The programmer can however, specify whether the class policy rule should be applied to sub-classes of the indicated class.
- *Smart Proxy Rules*, set using the *setFieldToBeCached()* method, permit individual fields of objects that are transmitted by-reference to be cached within proxies to those objects. If a field of a remotely accessible object is cached in a proxy for that object then the host holding the proxy can access the field value with the need for a network call.

An application programmer may specify or change policy rules at run-time, thus allowing for dynamic adaptation of the application. To specify policy rules statically, a library class programmer can specify the policy rules in the class' initialization code. The policy manager can also be configured to read and write policy rules stored in XML files, allowing the programmer to specify policies completely independently of the application source, as well as library class source.

Clearly, there is scope for contention between policy rules. For example, if an instance of class *X* is passed as a parameter to method *m*. A class policy rule may indicate that instances of *X* are passed-by-value while a method policy rule simultaneously indicates that parameters to method *m* are always passed-by-reference. As shown in Figure 7, each rule is specified as being overrideable or not. The RRT uses this information to break contention between rules by defining the following hierarchy in which the higher priority rules appear first:

1. Parameter policy rule (non-overridable)
2. Method policy rule (non-overridable)
3. Class policy rule (non-overridable)
4. Parameter policy rule (overridable)
5. Method policy rule (overridable)
6. Class policy rule (overridable)
7. Default policy

## Revisiting the Example

In our Peer-to-peer example introduced earlier, a *Message* might be transmitted by-value to an end-point using the *route* method on a *P2PNode*. However, if some of these objects are very large, the client programmer may with to transmit them by-reference. Figure 8 shows how the *deliver* method form Figure 6 may be modified to use the *TransmissionPolicyManager* to send those *Message* objects which exceed some maximum size by-reference, and smaller *Message* objects by-value.

```
public synchronized void deliver(Key destination
      IMessage message ) throws Exception {
   IP2PNode node = (IP2PNode)
         RAFDARunTime.getObjectByName(node, port, "P2P");
   if ( message.getSize() > MAX ){
      TransmissionPolicyManager.setClassPolicy(
           Message.class.getName(), BY_REF, true );
      node.route( destination, message );
   } else {
      TransmissionPolicyManager.setClassPolicy(
           Message.class.getName(), BY_VALUE, true );
      node.route( destination, message );
   }
}
```

**Figure 8: The modified deliver method**

Figure 9 illustrates the code necessary to instruct the *TransmissionPolicyManager* to make proxies for *P2PNodes* cache the immutable field *key* and that *Key* instances should always be passed by-value. On the client-side, the call to *getObjectByName* will yield a proxy of the remote *P2PNode* object which can be cast to the deployment interface type *IP2PNode*. A client holding such a proxy can access the *key* value of the remote *P2PNode* without having to make a remote call.

```
TransmissionPolicyManager.setClassPolicy(
      Key.class.getName(),BY_VALUE, true);
TransmissionPolicyManager.setFieldToBeCached(
      P2PNode.class.getName(), "key");
```

**Figure 9: Defining a smart proxy for P2PNode objects**

## Implementation issues

The deployment of an object requires several steps. Firstly a *skeleton* of the appropriate class is generated if necessary. Skeletons are the boundary between the application object (*servant* in Corba parlance) and the Web Services infrastructure. RRT *skeletons* all implement the interface shown in Figure 10. There is one *skeleton* class associated with each application class and one instance of a *skeleton* class is created and associated with each deployed object. Thus there is a one-to-one correspondence between *skeletons* and services. A *service map* maps from names and GUIDs to the skeleton associated with the particular service. The RRT automatically generates *skeleton* classes, instances of which reference the deployed objects and allow the RRT to perform method calls on them without using reflection. Skeleton generation incurs a one time cost and obviates the need for reflection during normal execution. Generated code is cached in the RRT for the duration of the JVM lifetime but can be configured to cache across multiple runs of the distributed application.

The *invokeMethod()* method allows the RRT to invoke a particular method with the supplied arguments on the underlying deployed object, while *getReturnType()* is used during automatic deployment to determine the signature types of exposed methods. The *getServiceObject()* and *init()* are used by the RRT to access the deployed object directly and to initialize the skeleton at instantiation time, respectively.

```
public interface SkeletonInterface {
   Object invokeMethod(String methodIdentifier,
   Object[] arguments) throws Exception;
   Class getReturnType(String methodIdentifier)
        throws Exception;
   Object getServiceObject();
   void init(Object serviceObject) throws Exception;
}
```

**Figure 10: The *Skeleton* Interface**

### Serialisation

During the object marshalling phase of a remote method call, the RRT will determine which object transmission semantics to employ and if pass by-value semantics have been chosen then it will serialize the closure of the return value. The serializer can handle the primitive SOAP types, such as *ints* and *strings*, by default and employs custom serializers to handle complex types. Custom serializers are singletons that are automatically generated on a per-class basis and each custom serializer is only capable of serializing instances of its associated application class.

All custom serializer classes provide two methods – one to serialize objects and another to perform deserialization. The *serialize()* method takes three arguments; the object to be serialized, the depth of this object within the closure of the return value being serialized, and an instance of the *SerializedObjects'* class.

### Implementing Remote References

The RRT implements remote references using remote identifiers, called *RRT Interoperable Object References* (RIORs), and proxy objects. RIORs uniquely identify deployed services in the distributed system and consist of:

- The machine name and port for the referenced object's RRT,
- Type information about the deployment interface used to create the service,
- The 128-bit randomly generated GUID,
- The programmer-defined service name,
- Smart proxy information.

To pass objects by-reference, the RRT serializes the associated RIORs by-value and, on deserialization, the client-side RRT uses it to initialize a proxy. Proxies, like skeletons and serializers, are automatically generated as required by the RRT and are created from the deployment interface type specified in the RIOR. If the deployment interface is a Java class then the proxy class extends it, while if it is a Java interface the proxy class implements it. As a result, the proxy is the same type as the deployment interface from the client's perspective. For every method in the deployment interface, the proxy implements an associated method with the same signature, which calls into the RRT to make a remote call to the deployed object on behalf of the client.

Application objects cannot make use of RIORs directly; they can only use references to other application objects or correctly typed proxy objects that have been initialized with the RIORs. Therefore, when RIORs are received by RRTs during remote method calls, the RRTs will convert them into references that the application can use. Initially, the RRT determines whether the referenced object exists in the local address space and if it does then a direct reference to the object is passed to the application. If not, the RRT determines whether a proxy to the referenced object has already been instantiated in the local address-space and, if the proxy exists then a reference to it is passed into the application. If a proxy does not already exist, then an instance of the associated proxy class is instantiated, automatically generating the class if necessary. This approach avoids the unnecessary use of remote references that loop-back into the same address spaces or the instantiation of more proxies than necessary.

### Smart proxies

All proxies have the capability to be smart proxies, which are proxies capable of caching some of the deployed objects' fields or code. RIORs contain smart proxy information indicating which fields and methods should be cached in the proxy and from this, an appropriate proxy class can be generated. The proxy class inherits the cached fields and methods from the deployment interface and the cached fields' *get()* and *set()* methods are modified to access the fields locally rather than invoke the equivalent method on the remote deployed object. Non-cached methods are overridden with proxy versions while cached methods are not overridden, leaving the original functionality in place. A new proxy class is generated for each combination of cached fields and methods in use within the distributed application.

Immediately before the RIOR is serialized, the RRT records the current values of the cached fields in it and they are serialized as part of the RIOR. On deserialization, the cached fields in the proxy object are initialized using a method similar to the custom serializer *init()* method described previously.

The RRT does not provide any form of automatic coherency control and so the programmer has responsibility for ensuring that application semantics remain as expected. Caching is particularly useful when object fields are known to be immutable.

### Custom Class Loaders

In order to implement proxies, custom serializers and skeletons, all applications class must be non-final and all their fields must be accessible to the RRT. Clearly not all classes written by application programmers comply with this requirement. A class loader is therefore provided that modifies application classes at load-time so that all fields are public and all classes are non-final. These transformations may not be made on classes in the standard Java libraries, resulting in the limitations with respect to the serialization of system classes described earlier.

### Automatic Deployment

The RRT can export references to un-deployed objects, for example, as return values or in the closure of returned objects. Automatic deployment ensures that appropriate deployment interfaces are chosen when exposing objects to remote access; the deployment interfaces must expose enough methods to preserve application semantics while not exposing any more methods than necessary to preserve the usefulness of the protection mechanism that the deployment interfaces provide. These two requirements are mutually antagonistic as the use of a deployment interface as a protection mechanism is by its very nature a restriction on the operations that can be performed on an object.

The process of automatic deployment proceeds as follows. If the object is deployed as a Web Service using its own class as a deployment interface, then no further deployment is required and the remote reference is typed as this service. If the object is deployed using other deployment interfaces, then if any of these are identical to or sub-types of the signature type, no further deployment is performed and the remote reference is typed as the narrowest of these types. Finally, if the object is not deployed using an interface that is related to the signature type or if it is not deployed at all, then the object is automatically deployed using its own concrete type as the deployment interface type.

The deployment of an object using the signature type preserves the protection mechanism role of the deployment interface but does not permit the client to cast the received object into a narrower type, even if such a cast is compatible with the actual type of the deployed object. The RRT can be configured to perform automatic deployment such that the object is always deployed using its own class as deployment interface. This approach means that all methods will be remotely accessible negating the protection mechanism but the proxy can be cast safely to any type compatible with the deployed object itself. The latter approach requires that private and protected methods as well as public methods be exposed in order to preserve local application semantics.

### Implementation of the Transmission Policy Framework

The policy framework is implemented using five associative stores, one for each rule type. Each associative store records argument policy rules and maps from keys to prioritized lists of policy rules. The keys are deterministically generated from the identity of the class and method being called and the argument numbers (where appropriate). To determine if an argument policy exists, the policy manager looks up the associative stores in order and if a mapping from the specified key exists, then the dominant argument policy rule is used. This approach is both simple and efficient.

The cost associated with evaluating the policy rules in order to determine which object transmission policy should be applied to a particular object is heavily dependent on the particular policy rules that are associated with the object to be transmitted. Figure 11 shows the cost imposed by the evaluation of policy on the overall remote method call time.

The test application performed a call on a remote method that took one argument and returned a return value. Both the argument and the return value were passed by-reference. The first row shows the time to perform one method call, averaged over 1600 method calls, without any policy evaluation phase, while the second shows the same set of calls with the policy evaluation phase included. In the latter case, the specified policy consists of a method policy rule and a return policy rule, both of which dictate that pass by-reference semantics should be employed. The test application represents the worst-case for the policy manager because no objects are serialized or transmitted and serialization of application objects increases the overall cost of the remote call and so proportionately decreases the cost of the policy evaluation phase. The introduction of additional arguments will have no effect on the proportionate cost of the policy evaluation phase.

| Time to perform 1 remote method call | Milliseconds |
|---|---|
| Without policy evaluation | 6.22 |
| With policy evaluation | 6.39 |

**Figure 11: Cost of policy evaluation on remote method invocation**

It can be seen that the policy evaluation phase has minimal impact on the overall cost of the remote method call—around 2% in this pathological case. In practice, the cost of dynamically evaluating policy is subsumed by the cost of marshalling and serialising the objects for remote method call. It is believed that the benefits gained outweigh the expense.

## Related Work

Web Services provide an RPC mechanism. A Web Service is a remote interface to a component class that has been deployed in a Web Service container. The Web Service container acts as a web server accepting incoming method calls in the form of HTTP requests. The URL specified in the request indicates which Web Service is being invoked. The body of the request contains the name of the method to invoke and the arguments to be passed, encoded using SOAP [8]. The Web Service Description Language (WSDL) [9] is used to describe the methods available in a Web Service.

Web Service technologies such as Apache Axis [6] and Microsoft .NET Web Services [4] deploy a class of component, not a specific instance of a component. The class is automatically instantiated to handle incoming requests on a per-call basis or on first access. Web Services systems do not permit the deployment of a specific component. Consequently, using standard Web Services, the only way in which specific components can be accessed is to manually provide a multiplexing Web Service which maps from keys to specific components. This makes it difficult to expose application components using standard Web Service technology.

Web Service technologies do not provide any form of remote object reference scheme. Web Services use only pass by-value semantics. In contrast, Distributed Object Models (DOMs) provide both RPC mechanisms and remote object reference schemes but do not allow arbitrary exposure of application components. A reference to a remotely accessible component can be passed across address space boundaries.

Method calls performed on the remotely referenced component are transparently propagated across the network to the referenced component.

The creation of a remotely accessible component using DOMs such as CORBA[1], Java RMI [2] and Microsoft .NET Remoting [4] requires the programmer to follow similar steps:

- The programmer is forced to decide statically the interfaces between distribution boundaries.
- The programmer is forced to decide statically which classes of component will implement these interfaces and thus be remotely accessible.
- These remotely accessible classes must extend a special base class that provides the functionality necessary for remote accessibility. This has two effects: to force the static identification of accessible classes, as above, and, in languages without multiple inheritance, to prevent the creation of accessible subclasses of existing non-accessible classes.
- Once a remotely accessible class is instantiated, the instance is associated with a naming service that allows remote callers to obtain a remote reference to it.

Some research DOMs, such as JavaParty [10] Fargo [11], and ProActive [12], have similar motivation to the work described in this paper and are briefly described below.

**JavaParty**

JavaParty [10] semantically extends Java with the addition of new keyword remote in order to simplify the process of creating remote classes. This keyword is permissible only in class signatures and indicates that instances of the class are remotely accessible. The JavaParty compiler generates pure Java code that uses RMI to implement remote accessibility. The generated Java and RMI source code is compiled in the usual manner to produce standard byte-code.

The remote keyword acts as a marker to the JavaParty compiler indicating which classes must be transformed into remote accessible versions. When creating a remote version of a class, the JavaParty compiler generates five Java classes, which replace the original class marked as remote. Initially, it separates the non-static and static members of the original class into two separate RMI enabled implementation classes, one of which contains only the non-static members and another which contains only the static members transformed into a non-static form. Accessor methods are generated for all fields and all members are made public so that interfaces can be extracted from each of these two classes. Extracted interfaces extend the RMI *java.rmi.Remote interface*. Finally, a wrapper class with the same name as the original class is generated.

This wrapper class holds interface-typed references to each of the generated implementation objects that capture the non-static and static functionality of the original class. However, as these implementation classes are RMI enabled the wrapper may actually be referencing RMI proxies to remote instances. Each method now acts as a wrapper method that calls its counterpart on the implementation object and handles any distribution related exceptions as best it can. JavaParty adopts the principle that this approach supersedes traditional RMI because an exception due to network failure is unlikely to occur on a local network, but if one does occur, it is unlikely that the programmer could handle it any better than the generated code. In addition, all RMI related code has been automatically generated negating the possibility of programmer-introduced errors at this level.

The motivation for JavaParty is similar to RAFDA. The major differences are in JavaParty's use of a pre-compiler and the integration of Web Services and Distributed Object Models along with the flexible policy framework provided by RAFDA.

**FarGo**

FarGo [11] implements an RMI-based DOM that supports migration and allows the programmer to impose policy rules on the references between objects. Like ProActive [12], the granularity of distribution is at the component level and the components are known as *complets*. A *complet* consists of a root object, known as an *anchor object*, and its closure, excluding any other anchor objects, which are considered the roots of distinct *complets*. Only the anchor object of a *complet* can be remotely referenced, though any object within a *complet* can hold a remote reference to a *complet* in another address-space. The infrastructure in which complets execute is known as a core, one of which exists in each address-space.

FarGo supports five types of remote reference:

- *link references* that are resilient in the face of *complet* migration, ensuring that referential integrity is preserved even if the referenced *complet* migrates to a new core.
- *pull references* express a migration policy between the reference holder and referenced *complet* indicating that if the reference holder migrates to a different core then the referenced complet should also migrate to that core.
- *duplicate references* indicate that if the reference holder migrates to a new core then it should take a duplicate copy of the referenced *complet* with it.
- *stamp references* indicate that after the reference holder migrates to a different it should rebind to any complet of the same class as the previously referenced *complet*.
- *bi-directional pull references* indicate if either the reference holder or referenced *complet* migrates to a different core then the other should also migrate to the same core,

The programmer expresses migration policy by reifying references in the application code and converting them into one of the above types or by specifying policy independently of source code using a scripting language.

**ProActive**

ProActive [12] is a Java library that provides tools for the creation of distributed applications using RMI or JMS for inter-address-space communication. Remotely accessible objects are known in ProActive as *active objects*, while all other objects are known as *passive objects*. Objects are made active by the programmer and one activated are remotely accessed and can be migrated from one address-space to another. Multiple active objects may not directly or indirectly reference a shared passive object. Each *active object* has a single thread executing within it that performs all computation on the Java objects. Method calls are queued up in each active object and serviced by this worker thread, which schedules and synchronises them.

Any non-final object can be made active by either instantiating it using a ProActive factory or by calling an activation method that takes an existing Java object and makes it active. The active version of an object comprises four Java objects; a conventional proxy and an object known as the body proxy are located in the client address-space and an object called the body proxy and the original object are located in the server address-space.

ProActive work is closest to the work described in this paper. It differs in that only certain *active objects* may be remotely accessed and sharing of passive objects is forbidden. In ProActive, each active object carries out the work performed by the RRT in our system. By contrast, in our work there is only one RRT instance per address space, which provides a view onto arbitrary application components. Furthermore, ProActive is based on RMI and JMS whereas the RRT is based on Web Services. Finally, ProActive does not provide the flexible transmission policy management supported by RAFDA.

### Java Management Extensions

Java Management Extensions [13] (JMX) provide a framework for the management and monitoring of Java applications. Resources are instrumented with Management Beans (MBeans) which must be implemented following specified design patterns. MBeans must specify or be associated with a Management Interface and it is only via this interface which external applications/clients may access the MBean. MBeans are registered with an MBean Server, responsible for mediating access to Mbeans to the MBean. The JMX framework supports access to the MBean Server via a variety of transport mechansims (RMI is standard), and allows for the implementation of custom transports. Related to the RRT, the framework includes mechansims to expose user defined classes (which may be precompiled) for management by an external application. In JMX, the management interface must be defined either statically (specifying and implementing a programmer defined interface) or dynamically by describing the Management Interface using standard Meta information classes. We have demonstrated that the facilities provided by the RRT may be used to provide the same functionality as JMX arguably but with more generality and less complexity.

## Conclusions

The RAFDA Run-Time (RRT) is a Java Middleware designed to improve the software engineering process for implementers of new distributed systems and for implementers of monitoring and management infrastructures aimed at existing applications. The work described in this paper has identified a number of key limitations exhibited by standard Middleware systems and had shown how the mechanisms provided by the RRT addresses each of these limitations.

Middleware systems typically require the programmer to decide at application design time which classes will support remote access and to follow a number of steps in order to create the remotely accessible classes. The programmer must decide the interfaces between distribution boundaries statically then determine which classes will implement these interfaces and thus be remotely accessible. This hard-coding of the distribution boundaries requires that the application programmer know if instances of a class will be remotely accessed before implementing that class.

The RRT permits instances of arbitrary classes within an application to be exposed for remote access. This is achieved through the dynamic deployment of a standard Web Service for the deployed object and the implementation of a mapping from remote calls on the Web Service to method calls on the deployed object. The RRT adds pass by-reference semantics to standard Web Services allowing methods on deployed objects to be called remotely.

In contrast to conventional Middleware systems, in order to deploy an instance of a class using the RRT, it is not necessary that class implement any special interfaces or extend any special classes. Thus the application programmer can implement the classes providing core application functionality without regard for the remote accessibility of the instances of those classes. Decisions about the remote accessibility of a particular object can be delayed until much later in the design cycle, even until run-time. Monitoring and management infrastructure that views and controls application state from another address space can be created without modification, or even access, to the application's original source code.

Another limitation of industry standard middleware systems is that the parameter-passing semantics is tightly bound to the distribution of the application and thus changes to the distribution of an application may potentially alter the application semantics.

The RRT addresses this limitation by providing a framework for the static and dynamic specification of object transmission policy. Using this framework the application programmer can employ the most advantageous object transmission policy for the particular circumstances. This increases flexibility and allows the programmer to control the application semantics. By specifying object transmission policy independently of class implementation, the roles of library class programmer and application programmer are separated. Library implementers must make less assumptions about the ways in which their classes will be used while application programmers can use class instances in the most appropriate way, as dictated by the particular situation. Before making a method call the application programmer can configure the transmission policy for the individual method parameters and any return value.

The transmission policy framework also supports the specification of smart proxies which increase the flexibility of deployed object without imposing implementation constraints on the programmer. This mechanism allows arbitrary field values of an object to be cached in the same address space as a remote reference (proxy) to that object. Thus a call to an accessor method on the proxy yields the field value without the execution of a network call.

The RRT employs dynamic code generation and compilation techniques to create the ancillary code necessary to allow dynamic object deployment. It is capable of marshalling instances of any class either by-reference or by-value and complete control over this is given to the programmer in order to separate parameter-passing semantics completely from application distribution.

The RRT provides significant advantages to programmers of distributed applications, when compared to industry standard Middleware systems, simplifying the software engineering process, decreasing the opportunity for errors in distribution code and increasing code reuse through better flexibility.

The RRT has been used in the construction of a global scale P2P routing network in which the application code can be run in both a fully distributed environment and in a centralised simulation environment without modification.